\documentstyle[preprint,aps]{revtex}
\draft

\begin{document}

\title{
Conserving and Gapless Approximations for an Inhomogeneous Bose Gas
at Finite Temperatures}

\author{A. Griffin}
\address{Department of Physics, University of Toronto,\\
         Toronto, Ontario, Canada, M5S 1A7\\  }

\date{\today}

\maketitle

\begin{abstract}
We derive and discuss the equations of motion for the condensate
and its fluctuations for a dilute, weakly
interacting
Bose gas in an external potential within the self--consistent
Hartree--Fock--Bogoliubov (HFB) approximation.  Account is taken of
the depletion of the condensate and the anomalous Bose
correlations, which are important at finite temperatures.  We give
a critical analysis of the self-consistent HFB approximation in terms of the
Hohenberg--Martin classification of approximations (conserving vs
gapless) and point out that the Popov approximation to the full HFB
gives a gapless single-particle spectrum at all temperatures.  
The Beliaev second-order approximation
is discussed as the spectrum generated by functional
differentiation of the HFB single--particle Green's function.
We emphasize that the problem of determining the excitation spectrum of a 
Bose-condensed gas (homogeneous or inhomogeneous) is difficult
because of the need to satisfy several different constraints.
\end{abstract}

\vspace{0.5cm}

\section{Introduction}
The recent observation\cite{and-etal95}   of Bose condensation in
 atomic gases trapped in an external potential well has
focussed attention on the excitation spectrum of an inhomogeneous
weakly interacting Bose-condensed gas at finite temperatures. 
Until very recently, this excitation spectrum was only addressed
indirectly in
the course of calculating \cite{gol-sil-leg81,hus-sig82} the local
density $n({\bf r})$\ and local condensate density $n_c({\bf r})$.

The excitation spectrum of a Bose
system is very dependent on the subtle dynamical correlations induced
by the Bose condensate \cite{hoh-mar65}.
In particular, one must calculate the single-particle and density-fluctuation 
spectrum in a consistent manner to ensure that they
are identical, as they must be
in a Bose-condensed system. In general, self-energy energy approximations
which lead to a single-particle spectrum without an energy gap
in the long-wavelength limit are not consistent with a density-fluctuation 
spectrum which satisfies conservation laws (or the corresponding sum rules).

A major purpose of this paper is to point out that
a useful way of understanding various approximations for the
excitations in a spatially non-uniform Bose gas is provided by the
kind of analysis developed in the early 1960's for a spatially uniform Bose
gas and codified in the well-known paper by
Hohenberg and Martin \cite{hoh-mar65}. It is based on distinguishing 
``conserving'' vs ``gapless'' approximations and the key role
played by 
the Hugenholtz-Pines theorem. 

We use this classification to examine
the equations of motion for  a non-uniform condensate and its
excited states in the finite temperature Hartree-Fock-Bogoliubov (HFB)
approximation, as well as simpler
approximations discussed in the literature.   An approximate
version of the HFB
due to Popov \cite{popov87,shi-ver-gri94} gives a simple gapless
approximation for the single-particle spectrum below the transition
(we do not discuss the critical region very near
$T_{BE}$)  but also reduces to the
$T=0$ Bogoliubov
approximation discussed by Fetter \cite{fetter92} for
a non-uniform gas. 
We also sketch the procedure \cite{hoh-mar65,che-gri71} for using
the 
HFB approximation for the single-particle self-energies to generate
by functional differentiation a density fluctuation spectrum which
is gapless in the long wavelength limit. This can be shown to be
identical to Beliaev's second order
approximation for the single--particle spectrum
\cite{beliaev58} at all temperatures. 

The emphasis in this paper is on the formal structure of various
kinds of approximations and how to assess their validity,
using the HFB approximation as a specific example. Our
results in (\ref{eq7}) and (\ref{eq22}) should provide a platform for future
numerical
calculations of the HFB excitation spectrum of an inhomogeneous Bose
gas at $T\not=0$, of the kind recently discussed by Edwards {\it et
al.}\ \cite{ref10} and Fetter \cite{newref11} using  the Bogoliubov
approximation at $T=0$ \cite{fetter92}. 
However, as we emphasize in the final section, the question
of what are the ``correct'' excitations in a dilute
weakly-interacting Bose gas is not simple one. It requires the full 
apparatus of many-body theory.

\section{Hartree-Fock-Bogoliubov Approximation}

Our starting point is the exact Heisenberg equation of motion 
\cite{fet-wal71,noz-pin90} for the Bose field operator
$\hat\psi({\bf r})$ 
\begin{equation}
i{\partial\hat\psi({\bf r}, t)\over\partial t} = \left
(-{\nabla^2\over
2m}+U_{ex}({\bf r}) -\mu\right )\hat\psi({\bf r}, t) +
g\hat\psi^\dagger
({\bf r},t)\hat\psi({\bf r}, t)\hat\psi({\bf r}, t)\ ,
\label{eq1}
\end{equation}
where we have assumed a short-range interaction $v({\bf r}-
{\bf r}^\prime)=g\delta({\bf r}-{\bf r}^\prime)$ between the
atoms (In the $s$-wave approximation, which is adequate for the
very dilute gases of interest, one has $g=4\pi a/m$). We assume the
atoms are trapped in a static
external potential $U_{ex}({\bf r})$.  Separating out the condensate part
in the usual fashion \cite{beliaev58,fet-wal71}, we have

\begin{equation}
\hat\psi({\bf r},t) = \Phi({\bf r}) +\tilde\psi({\bf r},t)\ ,
\label{eq2}
\end{equation}
where $\Phi({\bf r}) \equiv \langle\hat\psi({\bf r},t)\rangle=
\langle\hat\psi({\bf r})\rangle$ plays the role of a spatially-varying
macroscopic Bose field.  The thermal average in (\ref{eq2}) and
elsewhere is the usual anomalous average appropriate to a
Bose broken symmetry \cite{hoh-mar65}.  Using
(\ref{eq2}), the interaction term in (\ref{eq1}) can be written out in the form 
\begin{equation}
\hat\psi^{\dagger}\hat\psi\hat\psi =\vert\Phi\vert^2\Phi +
2\vert\Phi\vert^2\tilde\psi +
\Phi^2\tilde\psi^{\dagger} +\Phi^\ast\tilde\psi\tilde\psi +
2\Phi\tilde\psi^{\dagger}\tilde\psi
+\tilde\psi^{\dagger}\tilde\psi\tilde\psi\ .
\label{eq3}
\end{equation}

\noindent We treat the last term in (\ref{eq3}) in the
self-consistent
mean-field approximation, namely
\begin{equation}
\tilde\psi^{\dagger}({\bf r},t)\tilde\psi({\bf r},t)\tilde\psi({\bf
r},t)\simeq
2
\langle\tilde\psi^{\dagger}({\bf r})\tilde\psi({\bf
r})\rangle\tilde\psi({\bf r},t)+\langle\tilde\psi({\bf r})
\tilde\psi({\bf r})\rangle\tilde\psi^{\dagger}({\bf r},t)\ ,
\label{eq4}
\end{equation}
and then (\ref{eq3}) reduces to
\begin{eqnarray}
\lefteqn{\hat\psi^\dagger({\bf r},t)\hat\psi({\bf r},
t)\hat\psi({\bf
r}, t)=
\vert\Phi(r)\vert^2\Phi({\bf r})}\nonumber \\ 
&+& 2\left[\vert\Phi({\bf r})\vert^2 + \langle\tilde\psi^{\dagger}
({\bf r})\tilde\psi({\bf r})\rangle\right]\tilde\psi({\bf r},t)
+ \left[\Phi^2({\bf r})
+\langle\tilde\psi({\bf r})\tilde\psi({\bf
r}\rangle\right]\tilde\psi^{\dagger}({\bf r},t)\nonumber \\
&+ & 2\Phi({\bf r})\tilde\psi^{\dagger}({\bf r},t)\tilde\psi({\bf
r},t) +
\Phi^\ast({\bf r})\tilde\psi({\bf r},t)\tilde\psi({\bf r},t)
\ .
\label{eq5}
\end{eqnarray}

\noindent For the zero-range interaction we are considering, the
Hartree and Fock  (exchange) terms are
identical. This is the origin of the factor of 2 in (\ref{eq4}),
(\ref{eq5}) and subsequent equations.  

The time independent,
spatially-inhomogeneous Bose order parameter $\Phi({\bf r})$ is given
directly by taking an average over (\ref{eq1}),
\begin{equation}
\left(-{\nabla^2\over 2m} + U_{ex}({\bf r})-\mu\right)\Phi({\bf r})
+g\langle\hat\psi^\dagger({\bf r})\hat\psi({\bf r})\hat\psi({\bf
r})\rangle=0 \ .
\label{eq6}
\end{equation}
The time-dependence cancels out in the last term using the
cyclic invariance of the trace.
Taking the anomalous average of (\ref{eq5}), the linear terms in
$\tilde\psi({\bf r},t)$ vanish since
$\langle\tilde\psi({\bf r},t)\rangle=\langle\tilde\psi({\bf
r})\rangle=0$ and we  are left with
\begin{equation}
\langle\hat\psi^\dagger({\bf r})\hat\psi({\bf r})\hat\psi({\bf
r})\rangle =\vert\Phi({\bf r})\vert^2\Phi({\bf r}) + 2\Phi({\bf
r})\langle\tilde \psi^\dagger({\bf r})\tilde\psi({\bf r})\rangle
+\Phi^\ast({\bf r})\langle\tilde\psi({\bf r})\tilde\psi({\bf
r})\rangle
\label{neweq7}
\end{equation}
Using this in (\ref{eq6}), we find
\begin{equation}
\left(-{\nabla^2\over 2m} + U_{ex}({\bf r})-\mu\right)\Phi({\bf r})
+g[n_c({\bf r}) +2\tilde n({\bf r})]\Phi({\bf r}) + g\tilde m({\bf
r})\Phi^\ast({\bf r}) =
0 \ ,
\label{eq7}
\end{equation}
where we have introduced the local densities:
\begin{eqnarray}
n_c({\bf r})&\equiv&\vert\Phi({\bf r})\vert^2\nonumber\\
\tilde n({\bf r}) &\equiv &\langle\tilde\psi^\dagger({\bf
r})\tilde\psi({\bf r})\rangle\label{eq8}\\
\tilde m({\bf r})&\equiv&\langle\tilde\psi({\bf r})\tilde\psi({\bf
r})\rangle \ .\nonumber
\end{eqnarray}
\noindent We note that (\ref{eq7}) only reduces to a closed (non-linear) 
equation for $\Phi({\bf r})$ 
when both $\tilde n({\bf r})$ and $\tilde m({\bf r})$ are neglected
({\it i.e.}\ in the Bogoliubov approximation \cite{fetter92}). In
the case of a spatially-uniform Bose gas $(U_{ex}({\bf
r})\rightarrow 0)$, all the
functions in (\ref{eq8}) become constant, independent of position.

The excitations of the condensate are described by
$\tilde\psi({\bf r},t)$ in (\ref{eq2}) and
are given by the exact equation of motion
\begin{eqnarray}
i{\partial\tilde\psi({\bf r},t)\over \partial t} &=&
\left[-{\nabla^2\over 2m}
+ U_{ex}({\bf r})-\mu\right]\tilde\psi({\bf r},t)\nonumber\\
&&\mbox{\hspace{0.2cm}}+g\left[\hat\psi^\dagger({\bf
r},t)\hat\psi({\bf
r}, t)\hat\psi({\bf r}, t)-
\langle
\hat\psi^\dagger({\bf r})\hat\psi({\bf
r})\hat\psi({\bf r})\rangle\right]=0\ .
\label{eq9}
\end{eqnarray}
This follows from subtracting (\ref{eq6}) from (\ref{eq1}).
Consistent with our mean-field derivation of (\ref{eq7}),
the quadratic terms in the last line of
(\ref{eq5}) are treated in a mean-field approximation , {\it i.e.}
we
use
\begin{eqnarray}
\tilde\psi^\dagger({\bf r},t)\tilde\psi({\bf
r},t)&\simeq&\langle\tilde
\psi^\dagger({\bf r})\tilde\psi({\bf r})
\rangle\nonumber\\
\tilde\psi({\bf r},t)\tilde\psi({\bf
r},t)&\simeq&\langle\tilde\psi({\bf
r})\tilde\psi({\bf r})\rangle\ .
\label{eq11}
\end{eqnarray}
With (\ref{eq11}), the last term in (\ref{eq9}) is given by
\begin{eqnarray}
\lefteqn{
\hat\psi ^\dagger({\bf r},t)\hat\psi({\bf r}, t)\hat\psi({\bf r},
t) -
\langle\hat\psi^\dagger({\bf r})\hat\psi({\bf r})\hat\psi({\bf
r})\rangle}\nonumber\\
&&\mbox{\hspace{0.2cm}} \simeq
2\langle\hat\psi^\dagger({\bf r})\hat\psi({\bf
r})\rangle\tilde\psi({\bf r},t)
+\langle\hat\psi({\bf r})\hat\psi({\bf
r})\rangle\tilde\psi^\dagger({\bf r},t)\ .
\label{eq10}
\end{eqnarray}
Using (\ref{eq10}), (\ref{eq9}) reduces to 
\begin{eqnarray}
i{\partial\tilde\psi({\bf r},t)\over \partial t} &=& \left
(-{\nabla^2\over
2m}
+ U_{ex}({\bf r})-\mu\right )\tilde\psi({\bf r},t)\nonumber\\
&&\mbox{\hspace{0.5cm}}+2gn({\bf r})\tilde\psi({\bf r},t) +g
m({\bf r})\tilde\psi^\dagger({\bf r},t)
\ ,
\label{eq12}
\end{eqnarray}
where the self-consistent densities are defined by (see
(\ref{eq8}))
\begin{eqnarray}
n({\bf r})\equiv \langle\hat\psi^\dagger({\bf r})\hat\psi({\bf
r})\rangle
&=&\vert\Phi({\bf r})\vert^2
+\langle\tilde\psi^\dagger({\bf r})\tilde\psi({\bf
r})\rangle\nonumber\\
&\equiv&n_c({\bf r})+\tilde n({\bf r})
\label{eq13}
\end{eqnarray}
\begin{eqnarray}
m({\bf r})\equiv\langle\hat\psi({\bf r})\hat\psi({\bf r})\rangle
&=&\Phi^2({\bf r})
+\langle\tilde\psi({\bf r})\tilde\psi({\bf r})\rangle\nonumber\\
&\equiv&\Phi^2({\bf r}) +\tilde m({\bf r})\ .
\label{eq14}
\end{eqnarray}
One can easily derive the analogous equation of motion for
$\tilde\psi^\dagger({\bf r},t)$.

The coupled equations of motions given by (\ref{eq7}) and
(\ref{eq12})
correspond to the Hartree-Fock-Bogoliubov (HFB) approximation.  One
can derive the same or similar  results \cite{hus-sig82} by a
variety of ways but the above approach has the advantage when one
is interested in the excitations since it is easy to relate these
equations of motion to a more general Green's functions
\cite{hoh-mar65,che-gri71,beliaev58} formulation discussed in
Section~ III.

The preceding derivation of (\ref{eq7}) and (\ref{eq12}) is
equivalent to the grand canonical Hamiltonian
\begin{eqnarray}
\hat K\equiv \hat H -\mu\hat N &=&\int
d{\bf r}\hat\psi^\dagger({\bf r})\left[-{\nabla^2\over
2m}+U_{ex}({\bf
r})-\mu\right
]\hat\psi({\bf r})\nonumber\\
&+&{g\over 2}\int d{\bf r}\hat\psi^\dagger({\bf r})\hat\psi^\dagger({\bf
r})\hat\psi({\bf r})\hat\psi({\bf r}) 
\label{eq15}
\end{eqnarray}
being treated using a self-consistent quadratic approximation. 
Expanding the field operators in  this expression using
(\ref{eq2}), one finds after a
little algebra \cite{ref13}
\begin{eqnarray}
\hat K_{HFB}&=&\int d{\bf r}\Phi^\ast({\bf r})[\hat T({\bf r}) +
{1\over 2}
g n_c({\bf r})]\Phi({\bf r})\nonumber\\
&&+\int d{\bf r}\tilde\psi^\dagger({\bf r}) \hat{\cal
L}\tilde\psi({\bf
r})\nonumber\\
&&+{g\over 2} \int d {\bf r}
m({\bf r})\hat\psi^\dagger({\bf r})\hat\psi^\dagger({\bf
r})\nonumber\\
&&+{g\over 2} \int d {\bf r} m^\ast({\bf r})\tilde\psi({\bf
r})\tilde\psi({\bf r})\ ,
\label{eq16}
\end{eqnarray}
where $\hat{\cal L} \equiv \hat T ({\bf r})+ 2gn({\bf r})$, with
\begin{equation}
\hat T({\bf r})\equiv-{\nabla^2\over 2m} + U_{ex}({\bf r})-\mu\ .
\label{eq17}
\end{equation}
The coefficients of the linear terms in $\tilde\psi,\tilde\psi^\dagger$ in
(\ref{eq16}) can be shown to vanish by using 
the fact that $\Phi({\bf r})$ is given by the solution of
(\ref{eq7}). In going from (\ref{eq15}) to (\ref{eq16}), the {\it
only} approximation involves how we treat the terms cubic and
quartic in
$\tilde\psi$
and $\tilde\psi^\dagger$, namely (all quantities depend on ${\bf
r}$)
\begin{eqnarray}
\hat K_3 &\equiv&  g\int d{\bf r}
[\Phi^\ast\tilde\psi^\dagger\tilde\psi\tilde\psi
+\Phi\tilde\psi^\dagger\tilde\psi^\dagger\tilde\psi]\nonumber\\
&\simeq&g\int d{\bf r} [2\tilde n\Phi^\ast +\tilde
m^\ast\Phi]\tilde\psi+
g\int d{\bf r}[2\tilde n\Phi +\tilde
m\Phi^\ast]\tilde\psi^\dagger\label{eq18}\\
 \hat K_4&\equiv
&{g\over 2}\int d{\bf r}
\tilde\psi^\dagger\tilde\psi^\dagger\tilde\psi\tilde\psi\nonumber\\
&\simeq &{g\over 2}\int d{\bf r} [ 4\tilde
n\tilde\psi^\dagger\tilde\psi +\tilde
m^\ast \tilde\psi\tilde\psi +\tilde m
\tilde\psi^\dagger\tilde\psi^\dagger]\ .
\label{eq19}
\end{eqnarray}
That is, within the HFB  mean-field approximation, $\hat K_3$ is
linear in $\tilde
\psi,\tilde\psi^\dagger$
while $\hat K_4$ is quadratic in
$\tilde\psi,\tilde\psi^\dagger$.  We note that the coefficients in
(\ref{eq18}) and   (\ref{eq19}) involve either $\tilde n({\bf r})$
or $\tilde m$({\bf r}).

One can easily diagonalize
(\ref{eq16}) by using the linear
transformation
\begin{eqnarray}
\tilde\psi({\bf r})
&=&\sum_j(u_j({\bf r})\hat\alpha_j-v_j^\ast(r)\hat\alpha^
\dagger_j)
\nonumber\\
\tilde\psi^\dagger(r) &=&
\sum_j(u_j^\ast({\bf
r})\hat\alpha_j^\dagger-v_j(r)\hat\alpha_j) \
,
\label{eq20}
\end{eqnarray}
where $\hat\alpha_j$ and $\hat\alpha^\dagger_j$ are 
annihilation and creation operators satisfying the usual Bose
commutation relations. One can show that (\ref{eq16}) reduces to
\cite{ref14}
\begin{eqnarray}
\hat K_{HFB} &=&\int d{\bf r} \Phi^\ast({\bf r}) [~\hat T({\bf r})
+{1\over 2}
g\vert\Phi({\bf r})\vert^2]\Phi({\bf r})\nonumber\\
&& - \sum_j E_j\int d{\bf r}\vert
v_j({\bf r})\vert^2+\sum_jE_j\hat\alpha_j^\dagger\hat\alpha_j
\label{eq21}
\end{eqnarray}
if the $c$-number functions $u_j({\bf r})$ and $v_j({\bf r})$ are
given by
the solutions of the coupled eigenvalue equations
\begin{eqnarray}
\hat{\cal L} u_j({\bf r}) - gm({\bf r}) v_j({\bf r}) &=&
E_ju_j(r)\nonumber\\
\hat{\cal L} v_j({\bf r}) - gm^\ast({\bf r}) u_j({\bf r}) &=& -E_j
v_j(r)\ .
\label{eq22}
\end{eqnarray}
Here the operator $\hat{\cal L}$ is defined before (\ref{eq17}) and
$m({\bf r})$ is defined in (\ref{eq14}).  One can also
derive
the same results working directly with the equation of motion
(\ref{eq12}) and its Hermitian conjugate. In
particular, one may easily verify that
\begin{eqnarray}
\tilde\psi({\bf r},t) &=&
\sum_j[u_j({\bf r})\hat\alpha_je^{-iE_jt}-v_j^\ast({\bf
r})\hat\alpha
^\dagger_je^{iE_jt}]\nonumber\\
\tilde\psi^\dagger({\bf r},t) &=&
\sum_j[u_j^\ast(r)\hat\alpha_j^\dagger
e^{iE_jt}-v_j({\bf r})\hat\alpha_je^{-iE_jt}]
\label{eq23}
\end{eqnarray}
solves (\ref{eq12}) (and its adjoint) if $u_j, v_j$ and $E_j$
satisfy
the generalized ``Bogoliubov'' equations given in  (\ref{eq22}).
The results in (\ref{eq21}) and (\ref{eq22}) 
effectively reduce the problem to a
gas of non-interacting Bose quasiparticles with an energy spectrum
given by $E_j$.

We note that with (\ref{eq20}), one easily can derive expressions
for $\tilde n({\bf r})$ and $\tilde m({\bf r})$ in
(\protect\ref{eq8}) in
terms of
the self-consistent solutions of
the coupled equations (\ref{eq22}), namely
\begin{eqnarray}
\tilde n({\bf r}) &=&\sum_j\left\{\left[\vert u_i({\bf r})\vert^2
+\vert
v_j({\bf r})\vert^2\right]
N_0(E_j) + \vert v_j(r)\vert ^2\right\}\nonumber\\
\tilde m({\bf r}) &=&-\sum_j u_j({\bf r}) v^\ast_j({\bf r})
[2N_0(E_j) +1]\ ,
\label{eq24}
\end{eqnarray}
where $N_0(E)$ is the Bose distribution for the quasiparticle
excitations
\begin{equation}
{\langle\hat\alpha_j^\dagger  \hat\alpha_j\rangle}_{HFB} = {1\over
e^{\beta E_j}-1} \equiv N_0(E_j)\ .
\label{eq25}
\end{equation}
One must solve the coupled HFB equations (\ref{eq22}) and the
condensate equation (\ref{eq7}) using self-consistent values of
$\tilde n({\bf r})$ and $\tilde m({\bf r})$, given by (\ref{eq24}),
and $n_c({\bf r}) \equiv
\vert\Phi({\bf r})\vert
^2$.  Computationally, solving this set of coupled equations is
similar to
solving the
simpler $T=0$ Bogoliubov approximation
\cite{fetter92,ref10,newref11}. The
major difference is that (\ref{eq7}) is no longer a closed
equation
for $\Phi({\bf r})$ but involves the self-consistent diagonal and 
off-diagonal densities, $\tilde n({\bf r})$ and $\tilde m({\bf
r})$.  The
latter
depend on the excited states given by the solutions of the
self-consistent coupled
equations in (\ref{eq22}).

It is useful to summarize how one would solve the HFB equations
given by (\ref{eq7}) and (\ref{eq22}), for a given anisotropic
parabolic potential-well trap and temperature:
\begin{itemize}
\item[(a)]First calculate $n_0({\bf r})=|\Phi({\bf r})|^2$
and $\tilde{n}({\bf r})$ for a non-interacting gas ($g=0$). In this case, the
off-diagonal correlation function $\tilde{m}({\bf r})=0$.
\item[(b)]Use these non-interacting gas results in (\ref{eq22}) to solve
for $u_j$, $v_j$, and $E_j$. With these excited states, one can calculate
$\tilde{n}({\bf r})$ and $\tilde{m}({\bf r})$ from (\ref{eq24}) and use
these in (\ref{eq7}) to find $\Phi({\bf r})$.
\item[(c)]Iterate the above procedure to self-consistency.
\end{itemize}

Before discussing the implications and validity of (\ref{eq7}) and 
(\ref{eq12}), it is 
useful to relate these results to simpler approximations given in
the literature on inhomogeneous dilute Bose gases.  
We recall that (\ref{eq12}) and its adjoint are equivalent to
(\ref{eq22}) and (\ref{eq23}).  If we set both
$\tilde n({\bf r})$ and $\tilde m({\bf r})$ to zero, (\ref{eq7}) reduces
to the well-known Gross-Pitaevskii approximation, where $\Phi({\bf r})$
is given by a closed non-linear Schr\"{o}dinger equation (NLSE)
\cite{newref16}. Neglecting $\tilde{n}({\bf r})$ and $\tilde{m}({\bf r})$
in (\ref{eq12}) or (\ref{eq22}) leads to the standard
Bogoliubov approximation.  This simplified version of
Eqs.~(\ref{eq7}) and (\ref{eq12}) has been derived and discussed in
detail
by Fetter \cite{fetter92} 
(see also Refs.~\cite{ref10} and \cite{newref11}).  Since one is
ignoring the
non-condensate atoms $(\tilde n=0)$, this approximation is only
appropriate at $T\simeq 0$, where most of the atoms are in the
condensate $(n_c\simeq n)$.  Note that in the Bogoliubov
approximation, the cubic terms 
$(\hat K_3)$ and quartic terms $(\hat K_4)$ in (\protect\ref{eq18})
and (\ref{eq19}) are omitted completely.

The next level of approximating (\ref{eq7}) and (\ref{eq22})
would be to keep $\tilde n({\bf r})$ but neglect the anomalous density
$\tilde m({\bf r})$ as being small compared to both 
$\tilde{n}({\bf r})$ and $n_0({\bf r})$.  This approximation
has been used by Popov \cite{popov87} (in a homogeneous gas) to
discuss the finite-temperature region close to  the BE transition. 
This Popov approximation
\cite{shi-ver-gri94} formally reduces to the Bogoliubov
approximation at $T=0$, where $\tilde n$ also becomes negligible.
As we shall see in Section~III, it leads to a gapless spectrum,
in contrast with the full HFB.
Thus omitting $\tilde m({\bf r})$ but calculating $\tilde{n}({\bf r})$
in a self-consistent way seems to give a reasonable first approximation
for the excitation spectrum in Bose gases at
{\it all} temperatures.

The calculation of $\tilde n({\bf r})$ and $n_c({\bf r})$ given by Goldman
{\it et al.} \cite{gol-sil-leg81} involved equations equivalent to
(\ref{eq7}) with $\tilde m({\bf r}) = 0$ and to (\ref{eq22}) with 
$m({\bf r})=0$. 
This somewhat ad-hoc approximation is equivalent to putting
$v_j({\bf r})=0$ in (\ref{eq22}).
This simplified version of (\ref{eq22}) has been justified in 
Refs.~\cite{gol-sil-leg81,hus-sig82} 
as being adequate for the purpose of calculating 
$\tilde{n}({\bf r})$ and $n_0({\bf r})$ at temperatures just below $T_{BE}$,
even if it does not lead to a very good approximation for the excitation
spectrum.

\section{Conserving vs Gapless 
Approximations}

We now turn to a discussion of the HFB equations given by
(\ref{eq7}) and (\ref{eq12}).  The general problem of finding the
excitation frequencies in a homogeneous weakly interacting
Bose-condensed gas was  exhaustively studied 
and
various approximations were classified in a famous paper by
Hohenberg and Martin (see in particular Section VI of
Ref.~\cite{hoh-mar65}).  Any approximation put forward in the study
of an inhomogeneous Bose gas  can be usefully analyzed by examining
its
implications in the
corresponding homogeneous case.

For interacting Bose-condensed systems, it is convenient to
formulate the discussion in terms of Green's functions
\cite{hoh-mar65,fet-wal71}.  A key role is played by the $2\times
2$ matrix single-particle Green's function, defined as
\begin{equation}
G_1(1,1^\prime) = -i\langle
T\hat\Psi(1)\hat\Psi^\dagger(1^\prime)\rangle
\label{eq26}
\end{equation}
where
\begin{equation}
\hat\Psi(1)\equiv
{\pmatrix{\hat\psi(1)\cr \hat\psi^\dagger(1)}}\quad;\quad
\Psi^\dagger(1)\equiv (\hat\psi^\dagger(1),\psi(1))
\label{eq27}
\end{equation}
and 1 represents ${\bf r},t$.
Writing out the $2\times 2$ matrix in (\ref{eq26}) explicitly, we
have
\begin{equation}
G_{\alpha\beta}(1,1^\prime)=-\pmatrix{i\langle
T\hat\psi(1)\hat\psi^\dagger(1^\prime)\rangle & i\langle
T\hat\psi(1)\hat\psi(1^\prime)\rangle\cr
i\langle T\hat\psi^\dagger(1)\hat\psi^\dagger(1^\prime)\rangle &
i\langle T\hat\psi^\dagger(1)\hat\psi(1^\prime)\rangle}\ .
\label{eq28}
\end{equation}
Separating out the Bose condensate part of the field operator as in
(\ref{eq2}), (\ref{eq28}) naturally splits into two parts
\begin{equation}
G_{\alpha\beta}(1,1^\prime) =\tilde G_{\alpha\beta}(1,1^\prime) +
G_{\slantfrac{1}{2}\alpha}(1)
G^\dagger_{\slantfrac{1}{2}\beta}(1^\prime)\ ,
\label{eq29}
\end{equation}
where the condensate Green's function is described by
\begin{equation}
\hat G_{\slantfrac{1}{2}}(1) \equiv
\sqrt{-i}\langle\hat\Psi(1)\rangle
=\sqrt{-i}\pmatrix{\Phi(1)\cr \Phi^\ast(1)}
\label{eq30}
\end{equation}
\begin{equation}
\hat G^\dagger_{\slantfrac{1}{2}}(1)\equiv
\sqrt{-i}\langle\hat\Psi^\dagger(1)\rangle=\sqrt{-i}(\Phi^\ast(1)
,\Phi(1))\ ,
\label{eq31}
\end{equation}
and $\tilde G_{\alpha\beta}(1,1^\prime)$ is identical to
(\ref{eq28})
except that it involves the non-condensate part of the field
operators ($\tilde\psi$
and $\tilde\psi^\dagger$).  The signature of a Bose-condensed
interacting system is the appearance of anomalous correlation
functions such as $\Phi(1) = \langle\hat\psi(1)\rangle$ and $\tilde
G_{12} (1,1^\prime) = -i\langle
T\tilde\psi(1)\tilde\psi(1^\prime)\rangle$.

A very  convenient way of generating the equations of motion for
$\tilde G_1(1,1^\prime)$ and $\hat G_{\slantfrac{1}{2}}(1)$ is to
use functional differentiation with respect to external fields
\cite{hoh-mar65}. 
For homogeneous systems in thermal equilibrium, it is customary to
set these generating fields to zero at the end of the calculation.
However, in the context of the newly observed Bose-condensed gases
\cite{and-etal95}, these external fields are of direct physical
relevance, playing the role of the static external trapping
potential as well as time-dependent external perturbations
\cite{ref10,newref11}.  In this regard, the general formalism
developed in
Ref.~\cite{hoh-mar65} is particularly useful.  The general
equations of motion for $\tilde G_1(1,1^\prime)$ and $\hat
G_{\slantfrac{1}{2}}(1)$ are given by (2.6) and (2.7) of
Ref.~\cite{che-gri71}.  If we only include a {\it static} external
potential, the equation of motion for $\tilde G_1(1,1^\prime)$ is 
\begin{eqnarray}
\lefteqn{[i\tau^{(3)}_{\alpha\beta}{\partial\over \partial t} -
\hat T({\bf r})\delta_{\alpha\beta}]\tilde
G_{\beta\alpha^\prime}(1,1^\prime)}\nonumber\\
&&\mbox{\hspace{0.2cm}}=\delta(1-1^\prime)\delta_{\alpha\alpha^
\prime}
+\Sigma_{\alpha\beta}(1,\bar 2)\tilde G_{\beta\alpha^\prime}(\bar
2,1^\prime) \ ,
\label{eq32}
\end{eqnarray}
where $\tau^{(3)} = \pmatrix{1&0\cr
0&-1}$ and the single particle
operator $\hat T ({\bf r})$ is defined in (\ref{eq17}).  Here,
repeated Greek
indices are summed and a bar represents the usual integration over
${\bf r},t$.  This Dyson-Beliaev equations in (\ref{eq32}) for
the
single-particle Green's functions in a Bose-condensed system
depends on the $2\times 2$ matrix self-energy
$\Sigma_{\alpha\beta}(1,1^\prime)$ in addition to the chemical
potential $\mu$.
The analogous equation of motion for the two components of $\hat
G_{\slantfrac{1}{2}}$ is
given by (for a static external field)
\begin{equation}
-\hat T({\bf r}) G_{\slantfrac{1}{2}\alpha}({\bf r}) =
\sqrt{-i}\eta_\alpha({\bf r}) .
\label{eq33}
\end{equation}
Here we have introduced the condensate ``source function''
$\eta_\alpha({\bf r})$, which itself is a functional of $\hat G_{1/2}$ and
$\tilde G_1$. Writing (\ref{eq33}) more explicitly, it
corresponds to
\begin{equation}
\left[- {\nabla^2\over 2m} + U_{ex}({\bf r}) - \mu\right ]
\Phi({\bf
r}) = -
\eta_1({\bf r}) \ .
\label{eq34}
\end{equation}

One can now define what a ``conserving'' approximation is, namely
that there exists a functional $\Phi [\hat G_{\slantfrac{1}{2}},
\tilde G_1]$ of the correlation functions $\hat
G_{\slantfrac{1}{2}}$ and $\tilde G_1$ such
that the self-energy and the condensate source functions are given
by
\begin{eqnarray}
{\delta\Phi[\hat G_{\slantfrac{1}{2}}, \tilde G_1] \over \delta
\tilde
G_1(1, 1^\prime)} &=&\hat\Sigma (1, 1^\prime) \nonumber\\
{1\over 2\sqrt{-i}} {\delta\Phi[\hat G_{\slantfrac{1}{2}}, \tilde
G_1]
\over \delta \hat G_{\slantfrac{1}{2}}^\dagger (1)} &=&
\hat\eta(1)\ .
\label{eq35}
\end{eqnarray}
Martin and De Dominicis \cite{ref15} have proven that if such a
functional $\Phi$ exists, the two-particle Green's functions
generated by functional differentiation of $G_1(1,1^\prime)$ [given
by
(\ref{eq28}) - (\ref{eq31})] with respect to a time-dependent,
spatially-varying external field are
{\em guaranteed} to satisfy the usual conservation laws \cite{ref16}. 
The HFB approximation discussed in Section~ II can be shown to be an
example of such
a conserving approximation. Using Eq.~(6.42) of
Ref.~\cite{hoh-mar65}, one can easily verify that
\begin{equation}
\hat\Sigma_{HFB}(1,1^\prime) = g\pmatrix{
2n({\bf r}) & m({\bf r}) \cr m^\ast ({\bf r}) & 2n({\bf r})}
\delta(1-1^\prime)
\label{eq36}
\end{equation}
and
\begin{equation}
\hat\eta_{HFB}(1) = g\pmatrix{
n({\bf r})+\tilde n({\bf r}), & \tilde m({\bf r})\cr \tilde m^\ast
({\bf r}), & n({\bf r})+\tilde
n({\bf r})}
\pmatrix{\Phi({\bf r})\cr \Phi^\ast({\bf r})}\ .
\label{eq37}
\end{equation}
When substituted into (\ref{eq32}) and (\ref{eq33}), one finds
results equivalent to (\ref{eq22}) and (\ref{eq7}), respectively.
This HFB approximation is often called the Girardeau-Arnowitt
approximation in the Bose gas literature \cite{hoh-mar65}. Solving
(\ref{eq32}) using (\ref{eq36}), we arrive at the coupled
equations for $\tilde G_{11}$ and $\tilde G_{21}$,
\begin{eqnarray}
\left[ i{\partial \over \partial t} -\hat T({\bf r}) -
2gn({\bf r})\right ]\tilde G_{11}(1,1^\prime) - gm({\bf
r})\tilde
G_{21}(1,1^\prime) &=&\delta (1-1^\prime)\nonumber \\
\left[ -i{\partial \over \partial t} -\hat T({\bf r})-
2gn({\bf r})\right ]\tilde G_{21}(1,1^\prime) - gm^\ast({\bf
r})\tilde
G_{11}(1,1^\prime) &=&0\ .
\label{eq42}
\end{eqnarray}
This is the HFB generalization of the
Bogoliubov approximation, the latter being
given by (55.22) and (55.24) of Fetter and Walecka
\cite{fet-wal71} for inhomogeneous systems.
As expected, the structure of (\ref{eq42}) is identical to (\ref{eq22}).

The problem with a conserving approximation is that while the
two-particle Green's functions [such as the density response
function $\chi_{nn}(1,1^\prime)$] are guaranteed to have a spectrum
consistent
with conservation laws such as the equation of continuity
\cite{ref16}, the
single-particle
spectrum of $\tilde G_1(1,1^\prime)$ which one starts with may have
unphysical features.
In particular, one is not ensured of a gapless spectrum for long
wavelength excitations. In homogeneous systems, such a gapless
spectrum
of $\tilde G_1(1,1^\prime)$ is guaranteed if the self-energies
satisfy the Hugenholtz-Pines
(HP)
theorem, namely \cite{hoh-mar65,ref17}
\begin{equation}
\mu=\Sigma_{11}({\bf q} = 0, \omega=0) -\Sigma_{12}({\bf q}=0,
\omega =0)\ .
\label{eq38}
\end{equation}
More generally, it can be proven that this theorem will be
satisfied if the self-energies can be generated from the condensate
source function according to \cite{hoh-mar65}
\begin{equation}
\sqrt{-i} {\delta\hat\eta(1)\over \delta\hat
G_{\slantfrac{1}{2}}(1^\prime)} = \hat\Sigma(1,1^\prime)\ .
\label{eq39}
\end{equation}
Here $\hat\eta(1)$ is an explicit functional of $\hat
G_{\slantfrac{1}{2}}$ and implicitly through the dependence of
$\tilde G_1$ on $\hat G_{\slantfrac{1}{2}}$.  Approximations for
the self-energy consistent with (\ref{eq39}) are referred to as
``gapless'' approximations.  The HFB approximation for
$\hat\eta(1)$ and $\hat\Sigma(1,1^\prime)$ do {\it not} satisfy
(\ref{eq39}), and as a consequence the HFB spectrum of $\tilde
G_1(1,1^\prime)$ has a quasiparticle spectrum with a gap in the
limit of long wavelengths \cite{hoh-mar65}.  In contrast, the
simple
Bogoliubov approximation \cite{fet-wal71} is a gapless
approximation, as is the Beliaev second-order approximation
\cite{beliaev58}. Figs.~4 and 5 of Ref.~\cite{hoh-mar65} give a
convenient summary of conserving and gapless approximation for
uniform Bose gases.

We can illustrate the above somewhat formal discussion by
considering the HFB approximation for a uniform system
$(U_{ex}({\bf r})
\rightarrow 0)$. Using (\ref{eq37}) in (\ref{eq34}), or
equivalently (\ref{eq7}), we have (since $\Phi({\bf r}) =$
constant)
\begin{equation}
-\mu\Phi+g(n +\tilde n)\Phi +\tilde m\Phi^\ast = 0
\label{eq40}
\end{equation}
and hence the HFB gives the following result for the chemical
potential\cite{che-gri71}, 
\begin{equation}
\mu = g(n +\tilde n +\tilde m)\ .
\label{eq41}
\end{equation}

\noindent For a homogeneous
system, where we have  $\tilde
G_{\alpha\beta}
(1,1^\prime) =\tilde G_{\alpha\beta}(1-1^\prime)$, one can solve
(\ref{eq42}) by Fourier transformation. The poles of $\tilde
G_{\alpha\beta}({\bf k},\omega)$ are easily found to be  given by
\begin{equation}
E^2_k =\biggl({k^2\over 2m} -\mu +2gn\biggr ) ^2 - (gm)^2\ .
\label{eq43}
\end{equation}
Making use of the HFB chemical potential $\mu$ given by
(\ref{eq41}),
one sees that there is a finite energy gap at $k=0$, 
\begin{equation} E^2_{k=0} = g^2[n_c-\tilde m]^2 - g^2 [n_c +\tilde
m]^2=4g^2\vert\tilde m\vert n_c\ ,
\label{eq44}
\end{equation}
where we recall that $\tilde m$ is negative [see
(\ref{eq24})].  Clearly
the HFB does not satisfy the HP theorem (\ref{eq38}), since the
latter relation gives
\begin{equation}
\mu = 2gn -gm = g(n+\tilde n-\tilde m)\ ,
\label{eq45}
\end{equation}
which differs from the HFB result in (\ref{eq41}).  Using
(\ref{eq45}) in (\ref{eq43}) gives a gapless spectrum but this
procedure is ad hoc. One is using a chemical potential which is not
consistent with the equation of motion for
$\hat G_{\slantfrac{1}{2}}(1)$ or, equivalently, $\Phi({\bf r})$.

At the end of Section II, we introduced several approximate versions of 
the HFB results.  We can now discuss these approximations in terms
of the
excitation spectrum they lead to in the homogeneous case:
\begin{enumerate}
\item[(a)]``Bogoliubov'' corresponds to setting $\tilde m = \tilde
n = 0$.  Both (\ref{eq41}) and  
(\ref{eq45}) reduce to the same result $\mu=gn_c$ and hence the
single-particle excitations are gapless. This is valid at zero temperature.

\item[(b)]``Popov'' corresponds to setting $\tilde m = 0$ but
keeping $\tilde n$ finite.  Both (\ref{eq41}) and
(\ref{eq45}) reduce to $\mu=2gn-gn_c$ and hence the excitations are gapless.
As discussed at the end of Section~II, this gives a reasonable
approximation at all temperatures.

\item[(c)]In Goldman {\it et al.} \cite{gol-sil-leg81}, the
off-diagonal self-energies in (\ref{eq36}) involving $m({\bf r})$ are
neglected but only $\tilde m({\bf r})$ is neglected in the source term
(\ref{eq37}).  The result is that (\ref{eq43}) reduces to $E_k =
{k^2\over 2m} -\mu +2gn$ while (\ref{eq41}) reduces to $\mu =
gn+g\tilde n =2gn-gn_c$. Thus one obtains a spectrum $E_k = {k^2
\over 2m} + gn_c$ with an energy gap. This approximation can be
used for calculating $\tilde{n}$ and other thermodynamic quantities
near $T_{BE}$, where the
thermally significant excitations have a large enough wavevector
that the spectrum $E_k ={k^2 \over 2m} + gn_c$ is adequate
\cite{ref18}.
\end{enumerate}

In the Popov approximation to the HFB (in which the anomalous density
$\tilde m(r)$ is omitted), 
there is a special solution of the generalized Bogoliubov
equations (\ref{eq22}) corresponding
to $u_0({\bf r}) = v_0({\bf r})=\Phi({\bf r})$ with 
$E_0=0$,
where $\Phi({\bf r})$ is a solution of (\ref{eq7}) with $\tilde
m({\bf r})=0$.  As with the
simpler
Bogoliubov approximation discussed by Fetter \cite{fetter92,newref11}, one
sees in this case that the condensate wave function $\Phi({\bf r})$
plays the
role of the zero
energy single-particle mode and the fluctuations of the condensate
associated with $\tilde\psi$ are  described by
all the higher energy modes $(E_i>0)$.  In contrast, within the
full
HFB, there is no zero energy solution of (\ref{eq22}) corresponding
to
the solution $\Phi({\bf r})$ of (\ref{eq7}).

As we have noted above, conserving approximations for $\tilde
G_1(1,1^\prime)$ and $\hat G_{\slantfrac{1}{2}}(1)$ such as the HFB
can be used to generate two-particle Green's functions which are
guaranteed to satisfy conservation laws and related sum rules
\cite{hoh-mar65,ref15,ref16} and thus will lead to a gapless
density fluctuation spectrum.  In addition, we
recall that in the presence of a Bose broken symmetry, the
single--particle and density fluctuation correlation functions
exhibit the same poles \cite{ref17}.  This fundamental feature
holds for both uniform and non-uniform Bose systems, as can be
seen, for example, from Eq.~(6.38) of Hohenberg and Martin
\cite{hoh-mar65}.  Thus by using the HFB single--particle matrix
Green's function $G_1$ to generate 
the density response function by functional differentiation, one is
effectively generating an ``improved''
 single--particle spectrum which will be gapless. In particular,
as discussed
on p.~350--351 of Ref.~\cite{hoh-mar65}, the HFB $G_1$ generates a
density response function which is found to be identical to
Beliaev's second--order single--particle spectrum \cite{beliaev58}.  In the
uniform
case, this equivalence has been formally proven at {\it all}
temperatures in a Bose-condensed gas by Cheung and Griffin (see
Section IV of Ref.~\cite{che-gri71}).  The fact that this new spectrum
is gapless follows from the fact that the Beliaev self--energies
satisfy the HP relation in (\ref{eq38}), with $\mu$ being given by
the HFB result in (\ref{eq41}).

The above procedure shows that the excitations
given by the Beliaev second--order approximation is the correct
generalization on the HFB excitation spectra. Such calculations
involve evaluating  the various bubble
or polarization diagrams (see Eq.~(2.30) of
Ref.~\cite{che-gri71}) which involve products of two HFB
single--particle
propagators $\tilde G_1$.  However, at temperatures close enough to
$T_{BE}$, it is adequate to evaluate these polarization bubble
diagrams using a simple particle-like spectrum such as used in
Refs.~\cite{gol-sil-leg81,ref18}.

Within a variational calculation, Bijlsma and Stoof \cite{ref19}
have recently obtained (for a uniform system) the equivalent of the
Popov approximation as
defined above, but with $g$ replaced by the many-body $t$--matrix
calculated in the ladder diagram approximation.  This extension and
its relation to the full Beliaev second--order approximation will
be discussed elsewhere \cite{ref20}.

\section{conclusions}

Within the full Hartree-Fock-Bogoliubov approximation, we have
given a
simple derivation of the coupled equations of motion for the
condensate $\Phi({\bf r})$ and its excited states, as given by
(\ref{eq7}) and (\ref{eq22}) respectively.  In contrast with 
previous work which was variational in nature 
\cite{gol-sil-leg81,hus-sig82}, our
derivation in Section II works directly with a mean-field
approximation for the equation of motion for the condensate
$(\Phi)$ and non-condensate $(\tilde \psi)$ parts of the quantum
field operator. This brings out the physics involved most clearly
and has the advantage that one can make contact with the more
general Green's function formulation \cite{hoh-mar65,che-gri71}
discussed in Section III. The self-consistent HFB has, of course,
the advantage of being derivable variationally 
and thus the total energy is minimized even if the 
HFB excitations have certain deficiencies.

Because of the dynamical correlations induced
by the Bose-broken symmetry, a consistent theory of excitations is
surprisingly difficult to formulate even in a dilute Bose gas. 
This is already shown when one tries to improve the ``gapless''
Bogoliubov approximation by working with the ``conserving''
Hartree-Fock-Bogoliubov (HFB) approximation which is not
``gapless''. Hohenberg and Martin  (HM) \cite{hoh-mar65} give a
systematic discussion of this problem, introducing a
classification of different approximations.
In the present paper, we have emphasized the usefulness of the HM
classification when dealing with inhomogeneous atomic Bose gases in an
external potential well \cite{and-etal95}.  We have used it to
examine the single-particle spectrum predicted by the full HFB
\cite{hus-sig82} as well as various simplifications of it which have been
used in the literature \cite{gol-sil-leg81,popov87,fetter92}.
In addition, the HFB has a special significance since it can be
used to generate a density response
function by functional differentiation which turns out to have the 
same spectrum as the Beliaev second-order approximation\cite{beliaev58} for the
single-particle
Green's functions at all temperatures below $T_{BE}$.

The reason that finding the ``correct'' excitation spectrum
of a Bose-condensed gas is complicated is because
it is difficult to simultaneously satisfy different requirements within a given
approximation. 
The Hohenberg-Martin classification scheme we have used in Section~III of 
this paper does not
solve the problem of finding an approximation which is manifestly {\em both}
gapless and consistent with conservation laws. An alternative approach
which addresses this problem is based on the dielectric formalism
(for a review, see Section~5.1 of Ref.~\cite{ref17}). 
This is a diagrammatic procedure which builds in the close connection
between the single-particle and density-response functions
so that their excitation spectra are identical, with the conservation laws 
being incorporated through generalized Ward identities. This formalism 
\cite{ref23} cuts across the HM classification and gives a procedure
for choosing approximations which yield both a gapless elementary-excitation
spectrum and response functions consistent with conservation laws (sum rules).
In a future publication, we hope to discuss inhomogeneous Bose gases using 
this dielectric formalism approach \cite{ref23,ref17}.

\acknowledgments

I would like to thank Hua Shi for discussion as well as collaboration on parts 
of Section~II.  This work was supported by a research grant from NSERC.

\end{document}